\documentclass[pss,fleqn]{w-art}
\usepackage{times}
\usepackage{w-thm}
\usepackage{graphicx}
\begin{document}
\newcommand {\bea}{\begin{eqnarray}}
\newcommand {\eea}{\end{eqnarray}}
\newcommand {\nn}{\nonumber}
\newcommand {\bb}{\bibitem}
\DOIsuffix{theDOIsuffix}
\Volume{XX}
\Issue{1}
\Month{01}
\Year{2003}
\pagespan{1}{}
\Receiveddate{}
\Reviseddate{}
\Accepteddate{}
\Dateposted{}
\title{Bottom-up approach to high-temperature superconductivity}

\author[H. Won]{Hyekyung Won\inst{1}} 
\address[\inst{1}] {Department of Physics, Hallym 
University, Chuncheon 200-702, South Korea}
\author[S. Haas]{Stephan Haas\inst{2}}
\address[\inst{2}]{Department of Physics and Astronomy, University of Southern
California, Los Angeles, CA 90089-0484 USA}
\author[K. Maki]{Kazumi Maki\inst{2}}


\begin{abstract}
Since the discovery of high-temperature superconductivity in the cuprates
a theoretical understanding of their phase diagram has remained one of the
major outstanding problems in condensed matter physics. Here we propose an
effective low-energy Hamiltonian which produces both d-wave density wave (dDW)
and d-wave superconducting (dSC) solutions within the BCS mean-field theory. 
This model predicts that (a) the observed pseudogap phase is a dDW state,
(b) the superconducting phase is a d-wave BCS state, and (c) in the underdoped
regime there is a gossamer superconducting state, i.e. dSC in coexistence with
dDW. Moreover, this theory naturally explains
the Uemura relation, the reduction of
the quasiparticle density of states at the Fermi level, and the salient  
features in the tunneling conductivity measured in underdoped Bi2212. 
\end{abstract}
\maketitle                   

\section{Introduction}

In 1986, 
the discovery of high-temperature superconductivity in the cuprates took 
the physics community by surprise.\cite{1} This was the starting point of a 
new era in condensed matter physics. Many practical 
applications of these new compounds were envisioned, and at the same time 
an intense debate arose regarding the origin and possible mechanisms leading to 
this new phenomenon. 
Enz recorded the often confusing discussions of the early days of
high-Tc research in his beautiful textbook.\cite{2} One of the most influential 
contributions to the theory of these materials was provided by Anderson's
``dogmas".\cite{3} He stated that the cuprate high-Tc phase diagram
arises from an inherent competition between a Mott insulator phase and 
s-wave BCS superconductivity in these materials. 
In order to model high-Tc superconductivity, he proposed a two-dimensional 
one-band Hubbard model in combination with a resonant valence bond (RVB) 
wave function. A great portion of the theoretical community in the field 
has since embraced these dogmas. However, unfortunately we still remain 
without a clear vision as to where these dogmas are leading us.\cite{4} 
Around 1990, Scalapino and others\cite{5} pointed out that a perturbative 
analysis of the 2D Hubbard model in the weak-coupling limit produces d-wave 
superconductivity. Indeed, a d-wave superconducting order parameter was 
experimentally established around 1994 for single crystal samples of 
optimally doped Bi2212, YBCO and LSCO, using powerful angle resolved 
photoemission spectroscopy (ARPES)\cite{6} and elegant Josephson 
interferometry\cite{7,8}. These observations motivated us to investigate 
d-wave superconductivity within the BCS framework.\cite{9,10,11} 

\begin{figure}[h]
\includegraphics[width=8cm]{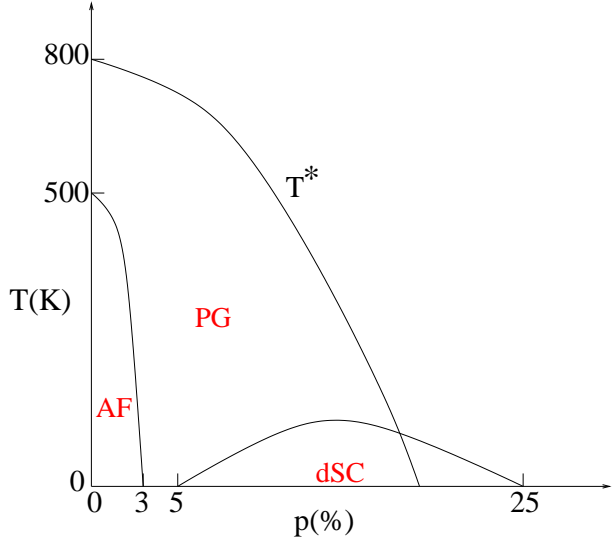}
\caption{The phase diagram for the high-T$_{c}$ cuprates. p denotes the hole 
doping concentration. PG is the pseudogap region.}
\end{figure}

Before elaborating further, let us first examine the generic phase diagram 
of the hole doped high-Tc cuprate superconductors shown in Fig. 1. From the
beginning, this phase diagram has been hotly debated. Around the year 2000,
a few groups suggested that the pseudogap region can be described by a 
d-wave density wave (dDW) phase. Indeed, the giant Nernst effect observed in 
the underdoped Bi2212, YBCO and LSCO\cite{17,18,19} and the angle dependent
magnetoresistance in Y$_{0.68}$Pr$_{0.32}$CuO$_4$ \cite{20} have been 
found to be fully consistent with dDW.\cite{21,22} In past work, we have shown 
that these are consequences of the Landau quantization of the quasiparticle
spectrum in a magnetic field, analogous to earlier 
considerations by Nersesyan et al.
\cite{23,24} Moreover, we note that the Fermi arcs (or pockets) in 
the ($\pi,\pi$) directions, observed by ARPES, follow directly from dDW.
\cite{16,25,26} Furthermore, it is by now well established that the overdoped 
regions of the cuprates 
can also be described in terms of a d-wave BCS model.
\cite{11,15}

Recently, Laughlin\cite{27}
pointed out that the Gutzwiller operator which is commonly 
used in the RVB wave function is not mathematically tractable, and proposed 
to replace it by a less constrained Jastrow operator. He named the resulting
coexistence phase
``gossamer superconductivity", i.e. a condensate with a reduced 
superfluid density and reduced density of states near the Fermi surface. 
Such a reduction of the quasiparticle density of states has recently been 
observed by Tallon et al\cite{28,29} by means of a thermodynamic analysis
and the effect of Zn impurities in YBCO over a wide doping range. In section 4
we will return to the characterization of gossamer superconductivity. 
Note also that gossamer superconductivity emerges naturally from the phase
diagram in Fig. 1 as a coexistent phase of d-wave superconductivity in the 
presence of a d-wave density wave.\cite{14,15,16} In the following we will 
explore this gossamer superconductivity phenomenon in detail. 

\section{Effective Hamiltonian} 

In this section,
we construct an effective low-energy Hamiltonian that constitutes
the basis of 
the bottom-up approach. This approach
should be viewed as an alternative
to the common top-down approaches originating from higher-energy 
Hamiltonians, such as the t-J and Hubbard 
models. Considering the energy scales of these models,
e.g. typically a Hubbard U of 
the order of 10$^6$K, it turns out to be a rather difficult task to arrive
at superconducting phenomena that exist at scales of Tc$\sim 10^3$K.
Based on the renormalization group analysis of 2D electron systems\cite{30} 
we understand that the normal state is a Fermi liquid, 
i.e. not a Luttinger liquid or bosonic liquid. Here we define the Fermi 
liquid via  a quasiparticle Green function which has simple poles, a definition
that is consistent with Shankar\cite{30} and Landau\cite{31}. 
Furthermore, we know 
that at sufficiently low temperatures the normal state becomes unstable against
infrared divergences in the 2-particle and/or 2-hole channels, implying 
superconductivity, or unstable against
divergences in the particle-hole channel, implying 
density wave phases. 

The effective low-energy Hamiltonian for such a system is given 
by\cite{14,32,33}
\bea
H &=&  \sum_{k,\sigma} \left( \epsilon_k - \mu \right) c^{\dagger}_{k,\sigma}
 c_{k,\sigma} -  \sum_{k,\sigma} \left( \Delta_1(k) c^{\dagger}_{k+Q,\sigma}
 c_{k,\sigma} + \Delta^*_1(k) c^{\dagger}_{k,\sigma} c_{k+Q,\sigma} \right)
\nn \\ 
 &-&  \sum_{k} \left(  \Delta_2(k) c^{\dagger}_{k,\uparrow} 
 c^{\dagger}_{-k,\downarrow} + \Delta^*_2(k) c_{-k,\downarrow}
 c_{k,\uparrow} \right) 
- g^{-1}_1 |\Delta_1(k)|^2 - g^{-1}_2 |\Delta_2(k)|^2,
\eea
where the amplitude and angular parts of the order parameters separate via
$\Delta_1(k) = \Delta_1 f(k)$ and $\Delta_2(k) = \Delta_2 f(k)$, and 
$Q \sim (\pi,\pi )$ is the nesting vector.
Two self-consistent gap equations follow directly from this Hamiltonian:
\bea
\Delta^*_1 &=& \frac{g_1}{\langle f^2(k)\rangle } \sum{k,\sigma} f(k) 
\langle c^{\dagger}_{k+Q,\sigma} c_{k,\sigma} \rangle ,
\\
\Delta^*_2 &=& \frac{g_2}{\langle f^2(k)\rangle } \sum{k,\sigma} f(k)
\langle c^{\dagger}_{k,\uparrow} c^{\dagger}_{-k,\downarrow} \rangle .
\eea
Here $\Delta_1$ and $\Delta_2$ are the order parameters of dDW and dSC 
respectively. In the following, 
we use $f(k) = \cos(2 \phi)$ as the angular 
dependence. 
A similar Hamiltonian has been considered in related work by 
Thalmeier.\cite{32} However this study was limited to conventional DW and
conventional SC, and to the case of vanishing chemical potential $\mu$. 
As we shall see here, $\mu$ is an important control parameter in the present 
model.\cite{16} Moreover, when both DW and SC are conventional, there 
is little room for their coexistence; instead phase separation is the rule.
\cite{32} On the other hand, when both order parameters are unconventional, 
there
is ample opportunity for their coexistence.\cite{33} This fact will be exploited in the following. 

The Nambu-Gorkov Green function\cite{34} of this model is given by 
\bea 
G^{-1}(k,\omega_n)=i \omega_n - \epsilon_k \rho_3 \sigma_3 + \mu \sigma_3
+ |\Delta_1|\exp(-i\phi_1 \rho_3 ) f(k) \rho_1 \sigma_3 +
|\Delta_2|\exp(-i\phi_2 \sigma_3 ) f(k)  \sigma_1,
\eea
and the corresponding spinor field is 
\bea 
\Psi_k = \left( c^{\dagger}_{k,\sigma}, c_{-k,-\sigma} , 
c^{\dagger}_{k+Q,\sigma} , c_{-k-Q,-\sigma} \right).
\eea
One notes that unlike in Ref. \cite{14}, the present $G^{-1}(k,\omega_n)$ 
possesses two Abelian gauge transformations associated with $\phi_1$ 
(sliding motion of dDW) and $\phi_2$ (supercurrent in dSC). The determinant 
of $G^{-1}(k,\omega_n)$ is given by 
\bea 
D = det| G^{-1}(k,\omega_n) | = \left(
\omega_n^2 + \epsilon_k^2 + \mu^2 + |\Delta_1(k)|^2 + |\Delta_2(k)|^2 \right)^2 
- 4\mu^2 \left( \epsilon_k^2 + |\Delta_1(k)|^2 \right).
\eea  
Using this result, the quasiparticle energy is found to be
\bea
E = \pm \sqrt{ \left( \sqrt{\epsilon_k^2 + |\Delta_1(k)|^2 } \mp \mu \right)^2
+ |\Delta_2(k)|^2 },
\eea
which agrees with earlier results.\cite{12,30}
Finally, the quasiparticle density of states is given by
\bea
\frac{N(E)}{N_0} =\left< \left|  Re \frac{E}
{\sqrt{ \left( \sqrt{\epsilon_k^2 - \Delta_2^2 f(k)^2 } \mp \mu \right)^2
- \Delta_1^2 f(k)^2}}
\left( 1 \mp \frac{\mu}{\sqrt{E^2 - \Delta_2^2 f(k)^2 }} \right)
\right| \right>,
\eea
where + and - stand for the positive-energy and negative-energy solutions
respectively. 

\begin{figure}[h!]
\includegraphics[width=8cm,angle=270]{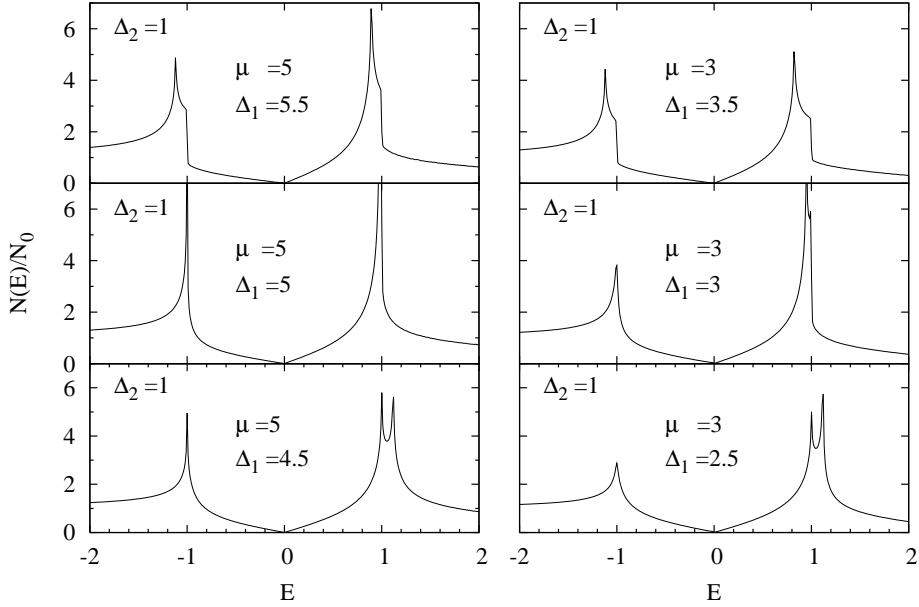}
\caption{Quasiparticle density of states in a gossamer superconductor 
with the energy scale set by the superconducting energy gap $\Delta_2 = 1$.}
\end{figure}

In Fig. 2 the quasiparticle density of states is shown for a particular 
set of parameters. For these parameters, we observe clear dips in the 
vicinity of E=0 , as well as a quasi-linear dependence on energy. 
In the regime $\Delta_1 - \mu  <\Delta_2$ one of the peaks splits into 
two peaks. Although such a split peak has not yet been observed
experimentally, except in 
the presence of Ni impurities,\cite{35} the present result is 
consistent with recent measurements on underdoped Bi2212.\cite{36} Moreoever,  
this result is rather different from earlier work by Zeyher and Greco \cite{37}
who assumed $\mu$ = 0. 

\section{D-Wave Density Wave Phase} 

Equations 2 and 3 can be transformed to\cite{33}
\bea
\lambda^{-1}_1 &=& 4\pi T \sum_n Re \langle f(k)^2 d^{-1} \rangle \\
\lambda^{-1}_2 &=& 4\pi T \sum_n Re \langle f(k)^2 Re \left[ \left(
1 - \frac{i \mu}{\sqrt{\omega_n^2 + \Delta_2^2 f(k)^2 }} \right) d^{-1} \right]
\rangle ,
\eea
where $ d = [ ( \sqrt{\omega_n^2 + \Delta_2^2 f(k)^2 } - \mu)^2 
+ \Delta_1^2 f(k)^2 ]^{1/2}$. Here $\lambda_1 = g_1 N_0$ and 
$\lambda_2 = g_2 N_0$ are dimensionless coupling constants, and $N_0 =
N(0)$ is the quasiparticle density of states in the normal state. 

First, the phase diagram of the pure dDW states
 is easily obtained by setting  $\Delta_2 =0$.
\cite{16} In this case, Eq. 10 reduces in the limit $\Delta_1 \rightarrow 0$ to
\bea
-ln \left( \frac{T_{c1}}{T_{c10}} \right) = Re \psi \left( \frac{1}{2} 
- \frac{i \mu}{2\pi T_{c1}} \right) - \psi \left( \frac{1}{2} \right),
\eea
where $T_{c1}$ is the transition temperature for dDW and $\psi (z)$ is the 
di-gamma function. Using $T_{c10}$=800K \cite{38}, one arrives at the phase
diagram shown in Fig. 3. Note that Eq. 13 is the same for s-wave and 
d-wave superconductors\cite{39,40,41} in the limit
when the Pauli term dominates over the orbital term.

\begin{figure}[h]
\includegraphics[width=8cm,angle=270]{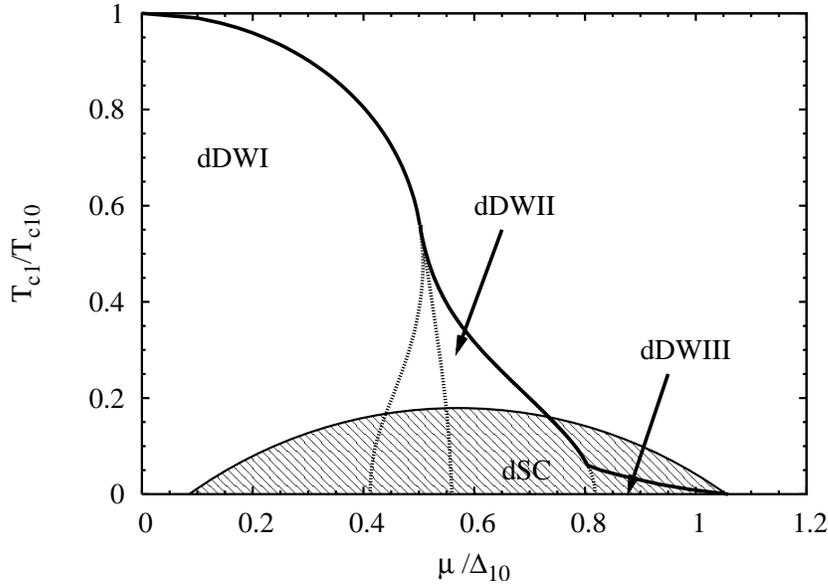}
\caption{The phase diagram of the high-T$_{c}$ cuprates. The shaded area
denotes $T_{c2}(\mu )$, whose determination from the present model will be 
presented in a future publication.}
\end{figure}

It is observed that $T_{c1}$ bends backwards in the region $\mu/\Delta_{10} 
\leq 0.558$. A similar diagram has also been found in Ref. \cite{37}. However, 
if we additionally allow a spatial variation of the dDW order parameter 
$\Delta_1$, we obtain
\bea
-ln \left( \frac{T_{c1}}{T_{c10}} \right) = Re 
\left< \left( 1 \pm \cos(4 \phi ) \right) \psi \left(\frac{1}{2} - 
\frac{i \mu (1 - p \cos(\phi)}{2\pi T_{c1}} 
\right) \right> - \psi \left(\frac{1}{2} \right),
\eea
where $p=v|q|/2\mu$.  This yields the extended portion shown in 
Fig. 3, analogous to the Fulde-Ferrell-Larkin-Ovchinnikov (FFLO) 
\cite{42,43} state in d-wave superconductors.\cite{44} There is a further 
transition when the q-vector is rotated from the [100] to the [110] direction. 
Finally, the dDW regime terminates when $\mu/ \Delta_{10} =0.824$. We 
call these 
periodic dDW phases dDWII and dDWIII respectively. 

\section{Gossamer Superconductivity}

Now we can ask how dSC appears on top of this dDW background. In the following, 
we shall limit ourselves to the region $\Delta_1 \gg \Delta_2, \mu$. Then we
can deduce in the vicinity of the superconducting transition temperature,
$T_{c2}$, the quasiparticle density of states at the Fermi surface which is 
given by 
\bea
g(\mu ,0) = \langle g (0,k) \rangle = \frac{2}{\pi} x K(x),
\eea
with $x = \mu/\Delta_1(\mu )$ and $K(z)$ is the complete elliptic function 
of the first kind. $g(\mu ,0)$ resembles the quasiparticle density of 
states deduced from the analysis of Zn-impurities\cite{28}. The corresponding 
low-temperature entropy is given by\cite{29}
\bea
\frac{S}{T} = \frac{27\zeta(3)}{2\pi^2} \gamma_N g(\mu ) \frac{T}{\Delta_2(\mu )},
\eea
where $\gamma_N = \pi^2 N_0/3$.
On the other hand, the superconducting transition temperature and free energy 
are controlled by
\bea
g_1 (\mu , 0 ) = 2 \langle \cos^2(2 \phi) g( 0, k) \rangle =
\frac{4x}{\pi} (K(x) - E(x)),
\eea
as
\bea
T_c(\mu ) &=& 1.136 \Delta_1 (\mu ) \exp[-(\lambda^{-1}_1 - \lambda^{-1}_2) 
g^{-1} (\mu , 0 )] , \\
U_0 &=& - \frac{1}{4} N_0 g_1 (\mu ) [\Delta_2 ( \mu, 0 )]^2.
\eea
Eq. (16) appears to somewhat 
overestimate $T_c(\mu )$, and hence a more detailed 
treatment needs to be developed in order to be more realistic. 
The functions $\Delta_1 (\mu)/\Delta_{10}$, $g(\mu , 0)$ and $g_1(\mu, 0) $ 
are shown in Fig. 4. The dependence of the
functions $g(\mu , 0)$ and $g_1(\mu, 0) $  on $\mu / \Delta_{10}$ is in
agreement with available experimental data.\cite{28,29} 

\begin{figure}[h]
\includegraphics[width=8cm,angle=270]{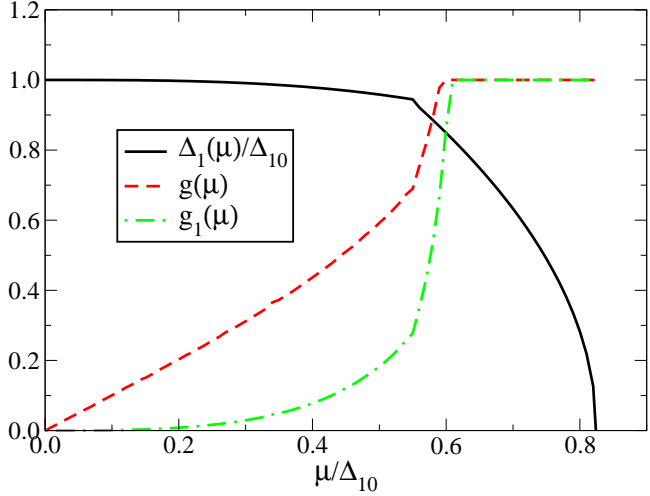}
\caption{Functions $\Delta_1 (\mu)/\Delta_{10}$, $g(\mu , 0)$ and 
$g_1(\mu, 0)$, 
controlling the low-doping region of the high-Tc superconductors. 
}
\end{figure}

By solving the coupled gap equations in the regime $\Delta_2, \mu \ll \Delta_1$ 
we find\cite{16}
\bea 
\rho_s (0) &=& \simeq \frac{\Delta_2^2 (\mu , 0 )}{\Delta^2 (0)} ,\\
T_{c2} &=&  \frac{1}{2 \ln 2} \frac{\Delta_2^2 (\mu , 0 )}{\Delta^2 (0)} ,\\
\lambda^{-2} (\mu, 0) &=&  \frac{4\pi e^2}{m^*} p \rho_s (0).
\eea
These are essentially the Uemura relations.\cite{45} 
Therefore, if we limit ourselves to the deeply underdoped region, 
many experimentally observed
features of gossamer superconductivity follow naturally
from the present model. 

\section{Concluding Remarks}

The model treated here is based on an effective low-energy Hamiltonian which 
describes dDW and dSC states
 with the chemical potential as a control parameter. 
This theory accounts for the following 
principle features of high-temperature cuprate
superconductors: (a) the normal state is a Fermi liquid, (b) the
pseudogap phase is a dDW (more recently a d-wave spin density wave has also 
been 
suggested\cite{46,47}), (c) the superconductivity in the optimal to overdoped 
regime has a BCS d-wave order parameter, and (d) in the underdoped regime 
there is gossamer superconductivity, i.e. dSC coexisting with dDW. 

Recent related studies on heavy fermion compounds, such as CeCoIn$_5$     
under pressure\cite{48}, and organic conductors, such as 
$\beta$''-(BEDT-TTF)$_4$[N$_3$O)M(C$_2$O$_4$]C$_5$H$_5$N with M = Ga and Cr
\cite{49,50}, have revealed many parallels between the high-Tc cuprates and 
these systems. These include (a) a layered structure or 
quasi-two-dimensionality, (b) d-wave superconductivity\cite{51,52}, and (c)
d-wave density wave phases\cite{47,53,54}. In the heavy fermion and 
organic conductors the horizontal axis in Fig. 3 needs to be replaced by 
the external pressure P, but otherwise their phase diagrams look very 
similar. This suggest strongly that the present model is rather universal and
applies to many strongly correlated electron systems. 

Let us finally
note that gossamer superconductivity is not necessarily restricted to 
dSC and dDW. For example, recent experiments on the Bechgaard salts
(TMTSF)$_2$PF$_6$ at ambient
pressure\cite{55,56,57} and measurements of the angle
dependent magnetothermal conductivity in URu$_2$Si$_2$\cite{58}
suggest that 
there are other kinds of gossamer superconductivity, e.g. 
f-wave superconductivity coexisting with a d-wave spin density wave. 
Hence it will not be surprising if similar coexistence states will soon be 
discovered in related strongly correlated electron systems.
                                                                                \begin{acknowledgement}
We are grateful for inspiring and stimulating discussions with Dionys
Baeriswyl, Christian Bernhard, Amalia Coldea, Balasz Dora, Jeff Tallon,
Peter Thalmeier, Andras Vanyolos and Attila Virosztek. K.M. and H.W. 
acknowledge the hospitality of the Max-Planck Institute f\"ur Physik 
komplexer Systeme at Dresden 
where most of this work was performed.
S.H. acknowledges financial support from the Petroleum Research Foundations, 
grant ACS PRF\# 41757 -AC10.
\end{acknowledgement}

\end{document}